\documentstyle[12pt]{article}

\newcommand\as{\alpha_{\mathrm{S}}}

\def\ep{\epsilon}
\def\ee{$e^+e^-$}
\def\beq{\begin{equation}}
\def\eeq{\end{equation}}
\def\beeq{\begin{eqnarray}}
\def\eeeq{\end{eqnarray}}
\def\cm{{\cal M}}

\def\bom#1{{\mbox{\boldmath $#1$}}}
\def\to{\rightarrow}

\def\RS{{\scriptscriptstyle\rm R\!.S\!.}}
\def\dsr{d\sigma_\RS^R}
\def\dsv{d\sigma_\RS^V}
\def\dsb{d\sigma_\RS^B}
\def\dsa{d\sigma_\RS^A}
\def\Trs{T^{\RS}}
\def\cmrs{{{\cal M}^{\RS}}}
\def\cf{{\cal F}}
\def\rs{regularization scheme}
\def\LO{leading order}
\def\NLO{next-to-leading order}

\def\gtil{{\tilde \gamma}}

\def\kt{k_\perp}
\def\CDR{{\rm CDR}}
\def\HV{{\rm HV}}
\def\DR{{\rm DR}}
\newcommand{\la}{\langle}
\newcommand{\ra}{\rangle}
\def\vspaceinarray{\nonumber ~&~&~\\}
\def\hg{h_g^\RS}
\def\P{\hat{P}^\RS}

\def\arrowlimit#1{\mathrel{\mathop{\longrightarrow}\limits_{#1}}}

\def\AP{Altarelli--Parisi }
\def\ID{1 \kern -.45 em 1}

\begin{document}

\begin{titlepage}
\renewcommand{\thefootnote}{\fnsymbol{footnote}}
\begin{flushright}
     CERN-TH/96-303 \\ hep-ph/9610553
     \end{flushright}
\par \vspace{10mm}
\begin{center}
{\Large \bf 
Regularization Scheme Independence and Unitarity \\[1ex]
in QCD Cross Sections }
\end{center}
\par \vspace{2mm}
\begin{center}
{\bf S. Catani}\\

\vspace{5mm}

{I.N.F.N., Sezione di Firenze}\\
{and Dipartimento
di Fisica, Universit\`a di Firenze}\\
{Largo E. Fermi 2, I-50125 Florence, Italy}

\vspace{5mm}

{\bf M.H. Seymour}\\

\vspace{5mm}

{Theory Division, CERN}\\
{CH-1211 Geneva 23, Switzerland}

\vspace{5mm}

{\bf Z. Tr\'{o}cs\'{a}nyi}\footnote{Zolt\'an Magyary fellow
of the Foundation for Hungarian Higher Education and Research.}\\

\vspace{5mm}

{Department of Theoretical Physics, KLTE}\\
{H-4010 Debrecen, PO Box 5, Hungary}

\end{center}

\par \vspace{2mm}
\begin{center} {\large \bf Abstract} \end{center}
\begin{quote}
\pretolerance 10000
  When calculating next-to-leading order QCD cross sections, divergences in
  intermediate steps of the calculation must be regularized.  The final
  result is independent of the regularization scheme used, provided that it
  is unitary.  In this paper we explore the relationship between
  regularization scheme independence and unitarity.  We show how the
  regularization scheme dependence can be isolated in simple universal
  components, and how unitarity can be guaranteed for any regularization
  prescription that can consistently be introduced in one-loop amplitudes.
  Finally, we show how to derive transition rules between
  different schemes without having to do any loop calculations.
\end{quote}

\vspace*{\fill}
\begin{flushleft}
     CERN-TH/96-303 \\   October 1996
\end{flushleft}
\end{titlepage}

\renewcommand{\thefootnote}{\fnsymbol{footnote}}

\section{Introduction}
\label{intro}

Two major bottlenecks hamper the straightforward
(but toilsome) implementation of QCD perturbation theory in physical
computations. The first that one encounters in producing new QCD
calculations for a certain process is the evaluation of the relevant 
matrix elements. The second regards the use 
of those matrix elements in the actual computation of 
physical quantities.
Recent years have witnessed much progress in perturbative QCD
calculations and both these major bottlenecks have been greatly reduced.
On the one hand, new techniques have been developed to 
evaluate QCD amplitudes at one loop [\ref{BDK96}] and on the other, 
completely general, process-independent methods [\ref{GGjets}--\ref{NTjets}]
have been set up to compute physical cross sections 
at \NLO\ (NLO). 

In particular,
reasonably compact expressions for all five-parton one-loop amplitudes
have been obtained [\ref{BDK5g}--\ref{BDK2q3g}] and NLO calculations 
for physical processes involving many partons have become feasible. 
The three-jet cross section in hadron collisions in the 
simplified case of pure-gluon subprocesses is now available
[\ref{pp3jetg},\ref{kilgore}], as is the four-jet cross section in
electron--positron annihilation in the simplified cases of leading colour
[\ref{4jleading}] and four-quark final states [\ref{4jqqqq}];
the full QCD results are expected to appear soon. 

These two bottlenecks are, however, strongly related, and
overcoming them independently is not sufficient for a successful implementation 
of perturbative QCD. The relationship is due to the regularization of 
unphysical divergences and, ultimately, to unitarity.
 
Theoretical evaluation of jet observables, i.e.\ infrared and collinear safe
quantities, should lead to unambiguous results for physical cross sections.
As a statement of principle this is almost trivial, but its practical
realization is far from trivial.  This is because expressions at
intermediate steps of the calculation (loop matrix elements and integrals
of tree-level matrix elements) contain ultraviolet, soft and collinear
divergences.
Consequently, one has to introduce some regularization procedure. Eventually,
the regularized singularities cancel in physical quantities and the finite
remainder should be independent of the regularization procedure. This
independence is only achieved, however, if the regularization prescription
is unitary.
The evaluation of the matrix elements and their implementation in the 
calculation of physical observables must therefore be carried out without
violating unitarity.

A regularization scheme (RS) that fulfils the requirement of
unitarity in QCD is known: conventional dimensional regularization. 
However, the intermediate ingredients that are necessary for a complete 
evaluation of cross sections are not always available within this
RS.
In fact, the new techniques for calculating one-loop QCD amplitudes use a
different version of dimensional regularization, dimensional reduction
[\ref{Sie79}]. 
Moreover, even when the squared
matrix elements in conventional dimensional regularization (or any other unitary
RS) are
available, calculations in other schemes may be simpler. Thus it is
worth while investigating whether the RS dependence can be controlled in a
simple way.

Within the context of dimensional regularization, the RS dependence of one-loop
amplitudes has been studied in detail. In Ref.~[\ref{KST2to2}],
from the general
structure of the 
squared matrix elements for all $2\to 2$ QCD
subprocesses, effective transition rules were derived to relate the NLO
loop corrections in conventional dimensional regularization to those in 
several different dimensional regularization schemes, including dimensional 
reduction. For practical purposes, this information on the matrix elements is 
sufficient for most of the QCD computations at NLO.

Using a mass regularization scheme, many subtleties related to a consistent
unitary implementation of QCD amplitudes in cross-section calculations 
have been pointed out in Ref.~[\ref{verm}] for the specific case of 
three-jet production in \ee\ annihilation. The treatment of more complicated 
processes along the lines of Ref.~[\ref{verm}] is far from trivial.

The general algorithm for computing jet cross sections presented in
Ref.~[\ref{CSdipole}] provides a very simple and transparent way to study
the RS independence and unitarity of QCD cross sections.  In this paper we
explore and illustrate some of the issues.  Our main results are:
\vspace{-2ex}
\begin{itemize}
\addtolength{\itemsep}{-1ex}
\item we show how the several RS-dependent ingredients can explicitly
be isolated in simple contributions to the 
NLO cross sections; these contributions are universal (they do not depend
on either the process or the jet quantity) and provide an explicit control
on the RS independence of the calculation;

\item in the general context of dimensional regularization, we 
derive the explicit RS
dependence of the one-loop QCD amplitudes 
in a simple way without doing any loop calculations; in particular, we confirm
the transition rules obtained in Ref.~[\ref{KST2to2}];

\item more generally, we can provide an explicit and simple
recipe to guarantee unitarity of cross-section calculations for any 
regularization prescription
that is consistently defined at the level of one-loop matrix elements.
\end{itemize}
\vspace{-2ex}

As for the last point, we should point out that the RS issue considered in
this paper does not regard the ultraviolet behaviour. Ultraviolet divergences
have to be properly regularized and then renormalized in off-shell Green
functions. This leads to the introduction of the running coupling $\as(\mu^2)$,
which we always assumed to be defined in a fixed renormalization scheme, say,
the ${\overline {\rm MS}}$ scheme (the renormalization-scheme dependence can 
always be controlled by an overall perturbative shift in $\as$).
The on-shell limit of the Green functions thus defines the singular (because
of soft and collinear divergences) loop amplitudes we are concerned with. 
These singularities are unphysical in the sense that they disappear in any
jet observable, i.e.\ in any physical quantity that is well-defined in QCD 
perturbation theory. Thus there is no need to consider only (soft and collinear)
regularization prescriptions that are completely justified on a field
theoretical
basis (for instance, to any perturbative order or for both tree-level and loop
amplitudes). 
One can introduce any 
regularization prescription that is
well-defined at the sole level of loop amplitudes: by explicit construction,
we shall show how one can then enforce unitarity in the cross-section
calculation at NLO.

The outline of the paper is as follows. In Sect.~2 we recall the general
and precise definitions of the NLO QCD cross sections we aim to calculate,
and discuss their relationship with the unitarity condition.
In Sect.~3 we recall the important features of a general
formalism --- the dipole formalism [\ref{CSdipole}] --- to calculate those
cross sections. We identify the terms that may contain RS dependence,
and write down the explicit conditions that ensure unitarity of physical
cross sections for any RS that is consistently defined at the level of
one-loop matrix elements.
In Sect.~4 we show that, within the general framework of dimensional
regularization, the explicit RS dependence of NLO QCD calculation can be derived
from that of corresponding the \AP splitting functions.  We do that for the
main schemes in current usage.
Sect.~5 contains our conclusions.

\section{QCD cross sections at \NLO\ }

According to the QCD factorization theorem, the hadron-level cross section
for a jet observable has the following expression 
\beq
\label{hadxs}
\sigma_{\rm had} = f \; \times  \; \sigma \; \times \; d \;\;,
\eeq
where $\sigma$ is the parton-level cross section and $f$ and $d$ are 
the non-perturbative parton densities 
and parton fragmentation functions of the incoming and observed 
(in the final state) hadrons, respectively.
The notation in Eq.~(\ref{hadxs})
is symbolic (see, for instance, Sect.~6 of Ref.~[\ref{CSdipole}] for a 
detailed notation). 
The hadronic cross section $\sigma_{\rm had}$ depends on the definition of the
jet quantity and on the
momenta of initial-state and final-state observed particles.
The parton-level cross section depends on the jet definition,
on the parton momenta and on the parton flavours.
The crosses in Eq.~(\ref{hadxs}) stand for the convolution over the momentum
fractions and for the sum over the flavours.

We recall that $\sigma_{\rm had}$ is a physical cross section while 
$f$, $d$ and $\sigma$ are not separately physical quantities. They depend on
the factorization scheme. Having defined this scheme, i.e.\ the
process-independent parton distributions $f$ and $d$,
the partonic cross section $\sigma$  is computable with no ambiguities
in perturbation theory to any order in $\as$. Its perturbative expansion up to 
NLO is the following:
\beq
\label{sig}
\sigma = \sigma^{LO} + \sigma^{NLO}.
\eeq
The \LO\ (LO) contribution $\sigma^{LO}$ is obtained by integrating the fully
exclusive cross section $d\sigma^{B}$ in the Born approximation over the phase
space for the corresponding jet quantity. Suppose that this LO
calculation involves $m$ partons in the final state. Thus, we write
\beq
\label{sLO}
\sigma^{LO} = \int_m d\sigma^{B} \;\;.
\eeq
Note that the LO cross section in Eq.~(\ref{sLO}) is finite by definition. 

Using analogous
notation, the NLO cross section $\sigma^{NLO}$ is a sum of three integrals:
\beq
\label{sNLO}
\sigma^{NLO} = \int_{m+1} d\sigma^R + \int_{m} d\sigma^V 
+ \int_{m} d\sigma^C \;\;.
\eeq
Here $d\sigma^C$ is a counterterm that defines the factorization scheme,
$d\sigma^R$ (the `real' cross section) is the exclusive cross section 
with $m+1$ partons in the final state, and $d\sigma^V$ (the `virtual'
cross section) is the one-loop correction to the process with $m$ 
final-state partons.

Strictly speaking, the expression on the right-hand side of Eq.~(\ref{sNLO})
has only a formal meaning because its contributions are separately divergent.
The virtual cross section $d\sigma^V$ is proportional
to the one-loop matrix element that, although renormalized,
still contains soft and collinear singularities coming from the loop 
integration. In order to remove these infinities, the loop integral has
to be regularized and, correspondingly, $d\sigma^V$ is replaced by its
(RS-dependent)
regularized version $\dsv$. The real cross section $d\sigma^R$ is finite
but its integration over the $m+1$-parton phase space produces soft and 
collinear divergences that cancel those in $d\sigma^V$, thus leading to a
finite NLO cross section\footnote{The integral of $d\sigma^R$ also produces
additional collinear divergences that cancel those in $d\sigma^C$.
Although these are in fact regularized by the same scheme as the others,
the resulting RS dependence can be absorbed into the factorization scheme
dependence of $d\sigma^C$, which does not concern us here.}
$\sigma^{NLO}$. This finite remainder is unambiguously
defined provided that, before its integration, the real cross section
$d\sigma^R$ is replaced by a consistently regularized version $\dsr$. A more 
correct way of writing Eq.~(\ref{sNLO}) is thus the following
\beq
\label{sNLOrs}
\sigma^{NLO} = 
\int_{m+1} \dsr + \int_{m} \dsv 
+ \int_{m} d\sigma^C \;\;,
\eeq
where some formal limit ${\rm R.S.} \to 0$ is understood.

Consistency of the RS means unitarity. More precisely, unitarity of the 
regularized theory implies that the scattering amplitudes $\Trs_{ab}$
that are used to compute its matrix elements have to fulfil the following
unitarity condition
\beq
\label{uni}
2 \,{\rm Im} \,\Trs_{aa} = \sum_b | \Trs_{ab} |^2 \;\;,
\eeq
up to the relevant order in perturbation theory. In particular, at NLO
the discontinuity of the one-loop matrix element (which is used
in $\dsv$) on the left-hand side of Eq.~(\ref{uni}) provides a constraint 
on the squares of the tree-level matrix elements (which are used in $\dsr$)
on the right-hand side.

The unitarity condition in Eq.~(\ref{uni}), which ultimately is at the basis of 
the cancellation theorems [\ref{cant}], also shows that the main
difficulty in controlling the cancellation of divergences in Eq.~(\ref{sNLO}) 
has a kinematic origin. If we had to compute the total cross section
for a given process, the real 
and virtual contributions could be combined at the integrand level before doing
the loop integral. In the soft and collinear regions, this integrand would 
contain exactly the difference between the left- and right-hand sides of 
Eq.~(\ref{uni}) and, hence, would be integrable even without introducing 
any regularization prescription. On the contrary, in Eq.~(\ref{sNLO}) the
integrations of the real and virtual contributions have to be performed over 
different phase-space regions. The shape of the two phase-space regions 
depends in a non-trivial way both on the number of
partons and on the actual definition of the cross section. In 
principle, one has to do a detailed calculation for any different jet observable
in any given process.

A general strategy [\ref{GGjets}--\ref{NTjets}] to overcome this
difficulty consists in trying to expose the cancellation of singularities
directly at the integrand level. This amounts to recasting Eq.~(\ref{sNLO}) in 
the following form:
\beq\label{sNLO4}
\sigma^{NLO} =
\sigma^{NLO\,\{m+1\}} + {\hat \sigma}^{NLO\,\{m\}} + \sigma^{NLO\,\{m\}} \;\;.
\eeq
In the contribution $\sigma^{NLO\,\{m+1\}}$, the
integration is carried out over the $m+1$-parton phase space.
The contributions $\sigma^{NLO\,\{m\}}$ and
${\hat \sigma}^{NLO\,\{m\}}$, instead, involve the
integration over the $m$-parton phase space. 

The precise definition of these three terms depends on the detailed method 
used in going
from Eq.~(\ref{sNLO}) to  Eq.~(\ref{sNLO4}). The common feature is that
the three integrands are separately finite. More precisely, the integrand
in $\sigma^{NLO\,\{m+1\}}$ is explicitly RS
independent
in the sense that it can be defined without introducing any \rs .
The integrand in ${\hat \sigma}^{NLO\,\{m\}}$ is also explicitly RS
independent, although it depends on the factorization scheme.
The integrand in $\sigma^{NLO\,\{m\}}$ contains the sum of two terms:
they are separately divergent if the regularization is removed. Both the
divergences and the RS
dependence (should) cancel in the  
sum. In the rest of this paper, we concentrate on the finiteness and RS
dependence of the integrand in $\sigma^{NLO\,\{m\}}$.

\section{Dipole formalism and the issue of unitarity}
\label{dipsec}

The dipole factorization formulae and the general algorithm for computing
QCD cross sections presented in Refs.~[\ref{CSLett},\ref{CSdipole}] 
are particularly 
convenient for studying the RS dependence of the integrand in 
$\sigma^{NLO\,\{m\}}$.
They indeed provide completely general and explicit expressions for the
cross-section contributions on the right-hand side of Eq.~(\ref{sNLO4}). 
In particular, all the kinematic complications related to the  
actual definition of the jet quantity are confined in overall 
factors.

The key point of the dipole formalism is the
universal definition of a `fake' cross section $\dsa$ that depends on the
momenta of the $m+1$ real partons involved in the evaluation of the NLO
cross section in Eq.~(\ref{sNLO}). The fake cross section has the following
general form
\beq
\label{dsA}
\dsa = \sum_{{\rm dipoles}} \;\dsb \otimes 
dV_{{\rm dipole}}^\RS \;\;.
\eeq 
In Eq.~(\ref{dsA}) the only dependence on the physical quantity
we are interested in is contained in the factor $\dsb$. This
is exactly the (regularized version of) Born-level cross section that enters in 
the calculation of the LO cross section $\sigma^{LO}$ in Eq.~(\ref{sLO}). The
only other ingredients needed to construct $\dsa$ are the dipole factors
$dV_{{\rm dipole}}^\RS$. They depend on the RS but are otherwise
universal: the dipole factors are completely process- and
observable-independent and can be given once and for all [\ref{CSdipole}] 
starting from the regularized $S$-matrix of the theory. The symbol $\otimes$
denotes properly defined correlations between colours and helicities of
the partons in $\dsb$ and in the dipole factors (see Sect.~\ref{cdrsec}).

The first main property of Eq.~(\ref{dsA}) is that, in the soft and collinear 
regions, $\dsa$ has the same pointwise singular behaviour as the (regularized)
real cross section $\dsr$ in Eq.~(\ref{sNLOrs}):
\beq
\label{ator}
\dsr \arrowlimit{{\rm soft\:and/or} \choose {\rm collinear}} \dsa \;\;.
\eeq
Equation (\ref{ator}) follows from the factorizing properties [\ref{BCM}]
of soft and collinear radiation in gauge theories and the 
dipole factorization theorem introduced in Ref.~[\ref{CSLett}]. The dipole
factors are precisely defined [\ref{CSdipole}] by the emission probability of
soft and collinear partons.

There are several dipole terms on the right-hand side of Eq.~(\ref{dsA}). 
Each of them corresponds to a different kinematic configuration of the $m+1$
real partons. Each configuration can be thought of as obtained by an effective
two-step process: an $m$-parton configuration is first produced and then one of 
these partons decays into two partons. The Born-level cross section $\dsb$
depends on the $m$-parton configuration, and the dipole factors describe
the one-to-two parton decays. This two-step pseudo-process can be defined 
without introducing any approximation on the $m+1$-parton kinematics,
thus leading to the second main property of Eq.~(\ref{dsA}): exact 
factorization of the phase space [\ref{CSdipole}]. 

Exact factorization means that we can carry out a factorizable
mapping from the $m+1$-parton phase space to an $m$-parton subspace,
identified by the partonic variables in $\dsb$, times a
single-parton phase space, identified by the dipole partonic variables in
$dV_{{\rm dipole}}^\RS$.
This mapping makes $dV_{{\rm dipole}}^\RS$ fully integrable
analytically. We can write:
\beq
\label{dsA1}
\int_{m+1} \dsa = \sum_{{\rm dipoles}} \;\int_m \;\dsb \otimes
\int_1 \;dV_{{\rm dipole}}^\RS = 
\int_m \left[ \dsb \otimes {\bom I}^\RS
\right] \;\;,
\eeq
where the universal factor ${\bom I}^\RS$ 
is defined by
\beq
\label{Ifac}
{\bom I}^\RS = \sum_{{\rm dipoles}} 
\;\int_1 \;dV_{{\rm dipole}}^\RS \;\;.
\eeq

These two main properties of $\dsa$ allow us to obtain the decomposition in
Eq.~(\ref{sNLO4})
by a straightforward implementation of the subtraction method [\ref{ERT}].
The fake cross section in Eq.~(\ref{dsA}) can be subtracted from $\dsr$ and 
$d\sigma^{C}$, and then added back to the right-hand side of 
Eq.~(\ref{sNLOrs}).
This subtraction defines the 
integrands of $\sigma^{NLO\,\{m+1\}}$ and
${\hat \sigma}^{NLO\,\{m\}}$ in Eq.~(\ref{sNLO4}), which, owing to
Eq.~(\ref{ator}), are explicitly RS independent, since the RS can be
removed already in the integrand.
The remaining term
defines $\sigma^{NLO\,\{m\}}$. Indeed, combining the virtual
cross section $\dsv$ with Eq.~(\ref{dsA1}), we can write this contribution 
as follows:
\beq
\label{sNLO3}
\sigma^{NLO\,\{m\}} = \int_m \left[ \dsv + \dsb \otimes 
{\bom I}^\RS
\right]_{\RS = 0} \;\;.
\eeq
The second term in the square bracket of Eq.~(\ref{sNLO3})
contains all the regularized singularities that are necessary to cancel 
the (equal and with opposite sign) regularized singularities in the virtual 
correction $\dsv$. After adding these two terms, one can thus remove
the regularization (as implied by the notation ${\rm R.S.} = 0$).

Actually, as explicitly denoted on the right-hand side of Eq.~(\ref{sNLO3}),
the regularization can be removed before carrying out the 
integration over the $m$-parton phase space. 
The reason for this is the following. Firstly, as recalled below 
Eq.~(\ref{sNLO}), soft and collinear divergences in $\dsv$ arise from the 
loop integral in the one-loop matrix element, independently of the definition
of the virtual cross section. Secondly, in the second term on the right-hand
side of Eq.~(\ref{sNLO3}) the dependence on the cross-section kinematics
is fully contained in $\dsb$, which, as pointed out below Eq.~(\ref{sLO}),
is integrable by definition. Following this observation, Eq.~(\ref{sNLO3}) can 
be rewritten so as to eliminate all the kinematic dependence from our 
discussion. 

The virtual cross section $\dsv$ can be written as follows:
\beq
\label{dsv1}
\dsv =  \sum_{\{ m \} } \, d\Phi^{(m)}(\{p\}) 
\;\;|\cmrs(\{p\})|^2_{{\rm 1-loop}} \;\;,
\eeq
where $\sum_{\{ m \}}$ stands for the sum over all the configurations with $m$ 
partons, $|\cmrs(\{p\})|^2_{{\rm 1-loop}}$ is the one-loop matrix element
squared and $\{p\}$ denotes its dependence on the parton momenta (these
include the $m$ final-state partons as well as possible partons in the initial 
state).
The factor $d\Phi^{(m)}(\{p\})$ in Eq.~(\ref{dsv1}) contains all the other
contributions to the differential cross section: 
spin and colour average factors, the $m$-parton phase space
and, in particular, the explicit definition
of the jet observable in terms of the parton momenta $\{p\}$.

The Born-level differential cross section $\dsb$, being itself defined on the 
$m$-parton phase space, has exactly the same form as Eq.~(\ref{dsv1}), apart
from replacing the one-loop matrix element $|\cmrs|^2_{{\rm 1-loop}}$ with the
corresponding tree-level  matrix element $|\cmrs|^2$. Thus, inserting 
Eq.~(\ref{dsv1}) and the analogous expression for $\dsb$ into Eq.~(\ref{sNLO3}),
we obtain
\beq
\label{sNLOf}
\sigma^{NLO\,\{m\}} = \int_m \;\sum_{\{ m \} } \, d\Phi^{(m)}(\{p\}) 
\;\;\cf(\{p\}) \;\;,
\eeq
where (omitting the notation $[ \dots ]_{\RS=0}$, from now on) 
we have introduced the following quantity 
\beq
\label{master}
\cf(\{p\}) = |\cmrs(\{p\})|^2_{{\rm 1-loop}} \;+  
\;|\cmrs(\{p\})|^2 \otimes 
{\bom I}^\RS(\{p\}) \;\;.
\eeq
The core of Eq.~(\ref{sNLOf}), namely the function $\cf(\{p\})$, embodies
all the relevant RS information. Although obtained by adding the two    
separately divergent and RS-dependent terms on the right-hand side of 
Eq.~(\ref{master}), this function is finite and RS independent in any 
unitary RS. 

Equation (\ref{master}) is our master equation for the study of 
unitarity and RS independence of physical cross sections. It can be considered
as the analogue of the unitarity condition (\ref{uni}). From a 
field-theory viewpoint, Eq.~(\ref{master}) is not as basic as Eq.~(\ref{uni}).
Nonetheless, it provides us with an explicit implementation of Eq.~(\ref{uni})
directly at the level of cross-section calculations and thus, it can be quite 
useful in practical terms. In order to discuss this point, let us consider the
main features of Eq.~(\ref{master}).

The master function $\cf(\{p\})$ is universal. It controls the soft
and collinear singularities of physical cross sections but depends
on process- and observable-independent contributions, the one-loop and 
tree-level matrix elements and the integral of the dipole factors. This makes
explicit the statement in the first item on the list in Sect.~\ref{intro}.

Within the context of the dipole formalism and the general algorithm for 
computing QCD cross sections of Ref.~[\ref{CSdipole}], Eq.~(\ref{master})
allows a straightforward implementation of different RSs in actual 
calculations of physical quantities.

In general, one can use Eq.~(\ref{master}) as a simple recipe to enforce and 
guarantee unitarity and RS independence in the evaluation of physical cross
sections. As recalled in Sect.~\ref{intro}, progress in the computation
of loop amplitudes cannot be disjoint from similar progress in the use of these
amplitudes for cross section calculations. In principle, one should regularize
unphysical soft and collinear divergences by introducing
a RS that is manifestly unitary, evaluate accordingly the real and virtual 
cross-section contributions in Eq.~(\ref{sNLO}), and perform all the steps
that are necessary to end up with finite physical cross sections.
Equation (\ref{master}) can be used as a convenient short cut
for this procedure. In order to actually evaluate  
$|\cmrs(\{p\})|^2_{{\rm 1-loop}}$, it is sufficient to introduce a 
regularization prescription that, at the level of one-loop amplitudes, is
defined in a consistent manner (it should not spoil general properties such
as gauge invariance and renormalizability). Then, one can compute the second
term on the right-hand side of Eq.~(\ref{master}) accordingly:
\begin{enumerate}
\item 
the partons in the tree-level matrix element $\cmrs(\{p\})$
have to be treated like the external partons in loop amplitudes;
 
\item
the parent parton and its (soft and collinear) decay partons
in the dipole factors $dV_{{\rm dipole}}^\RS$ have to be
treated like the partons inside loop integrals;

\item
the dipole phase space involved in the integral (\ref{Ifac})
of the dipole factors has to be treated like the phase space in the loops. 
\end{enumerate}
These rules are sufficient to calculate the master function $\cf(\{p\})$ that
contains all the relevant dynamical information on soft and collinear 
divergences. Having $\cf(\{p\})$ to hand, one can easily perform
RS transformations independently of the use of the dipole formalism in  
actual cross-section computations.

The reason why the unitarization recipe just outlined works is simple. The 
integral ${\bom I}^\RS(\{p\})$ of the dipole factors is proportional (but with 
opposite sign) to the discontinuity of the one-loop amplitude in the soft and
collinear regions. Therefore, any unitarity defect in the regularization
of the loop integral is automatically corrected by a corresponding
contribution in the dipole factors according
to Eq.~(\ref{uni}). Note that, in this respect, the concise notation in 
Eq.~(\ref{master}) may appear confusing. In fact, the tree-level amplitudes
$\cmrs(\{p\})$ are matrix elements of the regularized theory but do not
necessarily provide a complete set of them. The matrix elements  $\cmrs(\{p\})$  
are evaluated over the customary partonic states, while the saturation of
the unitarity condition (\ref{uni}) requires the sum over all possible
states that contribute to the discontinuity of the loop amplitude. The
regularization prescription of soft and collinear singularities in the loop
can introduce unphysical states\footnote{The role of these unphysical
  states in fulfilling the unitarity condition is somewhat analogous to
  that of the Faddeev--Popov (ultraviolet) ghosts.  In some specific cases
  they act as a negative number of scalar fields, so we can call them `infrared
  ghosts', although in general this analogy is only heuristic.  Because of the
  way in which we treat them,
  they are more like Feynman's `dopey particle' [\ref{dope}] than the
  formal ghosts of Faddeev and Popov.}: in Eq.~(\ref{master}) the
contribution of these is cancelled by analogous terms in the integral of
the dipole factors.

Note also that our unitarization recipe does not require the actual calculation
of loops. It is sufficient to integrate the dipole factors, which 
are tree-level objects in every respect. This is a non-trivial computational
simplification.

The use of infrared regularization prescriptions that are not manifestly 
unitary may appear an oddity. However, there are in fact such schemes
in current practice.
In the case of ultraviolet divergences, for instance, an unequal treatment
of particles inside the loop and external particles can easily be reconciled
with unitarity. A similar unequal treatment in the case of soft and collinear
divergences is not so harmless. In the calculation of physical cross sections,
the infrared singularities of loop amplitudes are cancelled by
corresponding singularities arising from the integration of tree-level matrix
elements: in the latter, it is not so trivial to make a distinction between 
`internal' and `external' particles. This point can be clarified by the
explicit examples considered in Sect.~\ref{dimregsec} in the context of 
dimensional-regularization prescriptions.
  
\section{Dimensional regularization}
\label{dimregsec}

\subsection{Conventional dimensional regularization}
\label{cdrsec}

The RS known as conventional dimensional regularization 
simultaneously regularizes ultraviolet [\ref{tHV},\ref{bollini}] and
soft and collinear divergences [\ref{cdrir}]. It amounts to 
analytically continuing parton momenta to $d=4-2\ep$ space-time dimensions and 
to considering $d-2$ helicity states for gluons and 2
helicity states for massless quarks. No distinction is made between real
and virtual partons. This RS is manifestly Lorentz and gauge invariant and
consistent with unitarity. 

In Ref.~[\ref{CSdipole}], the calculations were carried out using
conventional dimensional regularization. Here we summarize the explicit
results. For this purpose, we recall some notation.
 
It is useful to introduce a basis 
$\{ |c_1,\dots,c_n\ra \otimes |s_1,\dots,s_n\ra \}$ in colour + helicity space
in such a way that the tree-level matrix element with $n$ partons can be written
as follows:
\beq
\label{medef}
\cm^{c_1,\dots,c_n;s_1,\dots,s_n}(p_1,\dots,p_n) \equiv
\left( \frac{}{}\la c_1,\dots,c_n| \otimes \la s_1,\dots,s_n| \;\right) \;
|1,\ldots,n\ra \;\;,
\eeq
where $\{c_1,\dots,c_n\}$, $\{s_1,\dots,s_n\}$ and $\{p_1,\dots,p_n\}$ are 
respectively colour indices, spin indices and momenta of the partons. Thus
$|1,\ldots,n\ra$ is a vector in colour + helicity 
space\footnote{In the case of initial-state partons, the definition of the 
state vector $|1,\ldots,n\ra$  in Eq.~(\ref{medef}) differs by a
normalization factor
(proportional to the number of colours) with respect to the definition used
in Ref.~[\ref{CSdipole}] (cf. Eq.~(3.11) in [\ref{CSdipole}]).}.

According to this notation, the (RS-dependent) tree-level matrix element squared
summed over colours and helicities is:
\beq
|\cmrs(\{p\})|^2 = 
{}_{\RS}\la 1,\ldots,n | 1,\ldots,n \ra_{\RS} \;\;,
\eeq
while the colour and spin correlations denoted by $\otimes$ in 
Eq.~(\ref{master}) (and everywhere throughout the paper) are given as follows:
\beq
\label{insop}
|\cmrs(\{p\})|^2 \otimes {\bom I}^{\RS}(\{p\}) =
{}_{\RS}\la 1,\ldots,n | \;{\bom I}^{\RS}(\{p\})\; 
| 1,\ldots,n \ra_{\RS} \;\;,
\eeq
where the integral ${\bom I}^{\RS}$ of the dipole factors is a matrix in 
colour + helicity space and acts as an insertion operator on the right-hand
side of Eq.~(\ref{insop}).

In conventional dimensional regularization, although the dipole factors are 
helicity-dependent, spin correlations vanish after integration over the dipole
phase space. Thus, the insertion operator\footnote{Since we consider 
conventional
dimensional regularization as the default case, we drop the label RS in this 
scheme.} ${\bom I}(\{p\},\ep)$ is diagonal in the helicity space and depends
only on the colour charges ${\bom T}_I$ (see Sect.~3.2 of [\ref{CSdipole}]
for their detailed definition) and momenta $p_I$ of the partons in 
the tree-level matrix element $\cm(\{p\})$ $(I=1,\dots,n)$. Its explicit 
expression is [\ref{CSdipole}]:
\beq
\label{iee}
{\bom I}(\{p\},\ep) = -
\frac{\as}{2\pi}
\frac{1}{\Gamma(1-\ep)} \sum_I \;\frac{1}{{\bom T}_{I}^2} \;{\cal
V}_I(\ep)
\; \sum_{J \neq I} {\bom T}_I \cdot {\bom T}_J
\; \left( \frac{4\pi \mu^2}{2 p_I\cdot p_J} \right)^{\ep} \;\;,
\eeq
where $\mu$ is the dimensional-regularization scale and the singular (for
$\ep \to 0$) function ${\cal V}_I(\ep)$ depends only on the parton flavour and 
has the following $\ep$-expansion:
\beq
\label{calvexp}
{\cal V}_{I}(\ep) = {\bom T}_{I}^2 \left( \frac{1}{\ep^2} -
\frac{\pi^2}{3} \right) + \gamma_I \;\frac{1}{\ep}
+ \gamma_I + K_I + {\cal O}(\ep) \;\;.
\eeq
For present purposes there is no need to recall the detailed 
calculations leading to Eqs.~(\ref{iee},\ref{calvexp}). It is sufficient to note
that the constants $\gamma_I$ and the $K_I$ in Eq.~(\ref{calvexp}) 
are related to the $d$-dimensional integral of the \AP splitting functions
consistently evaluated in conventional dimensional regularization 
(cf. Sect.~\ref{apsec}). As a matter of fact, we have
\beq
\label{coef}
- \frac{1}{2} \sum_b \int_0^1 dz \;\left( z(1-z) \right)^{-\ep} \;
\la {\hat P}_{ab}(z;\ep)\ra = 2 {\bom T}_a^2 \frac{1}{\ep} + \gamma_a
+ \left( K_a - \frac{\pi^2}{6} {\bom T}_a^2 \right) \ep + 
{\cal O}(\ep^2) \;\;,
\eeq
where $\la {\hat P}_{ab}(z;\ep)\ra$ denotes the {\em azimuthally averaged} 
splitting function.

For the sake of completeness, the actual values of the constants entering into 
Eq.~(\ref{calvexp}) are:
\beeq
{\bom T}_q^2 = {\bom T}_{\bar q}^2 = C_F  \;\;, 
\;\;\;\;\;\;&&{\bom T}_g^2 = C_A \;\;, \nonumber \\
\gamma_q = \gamma_{\bar q} = \frac{3}{2} \,C_F \;\;, 
\;\;\;\;\;\;&&\gamma_g = \frac{11}{6} \, C_A -  \frac{2}{3} \,T_R N_f \;\;, \\
K_q = K_{\bar q} = \left( \frac{7}{2} - \frac{\pi^2}{6} \right )\,C_F \;\;, 
\;\;\;\;\;\;&&K_g = \left( \frac{67}{18} - \frac{\pi^2}{6} \right ) \, C_A 
-  \frac{10}{9} \,T_R N_f \;\;,\nonumber
\eeeq
where $C_F=(N_c^2-1)/2N_c, \,C_A=N_c, \,T_R=1/2$, $N_c$ is the number of colours
and $N_f$ is the number of massless flavours.

\subsection[]{Regularization prescriptions within\\dimensional regularization}
\label{rssec}

Dimensional regularization was invented to regularize ultraviolet
divergences in loop integrals of gauge theories
in a gauge-invariant manner. In this respect
the essential ingredient is the continuation of loop
momenta into $d\ne 4$ dimensions. Having done this, one is left with some
freedom regarding the dimensionality of the momenta of the external
particles as well as the number of polarizations of both external and
internal particles. 

The original choice of 't~Hooft and Veltman [\ref{tHV}] was to
continue the particle momenta and the helicities of vector particles inside 
loops into
$d\ne 4$ dimensions, while keeping the momenta and helicities of external
particles, as well as fermion helicities inside loops, in four dimensions.

Another version of dimensional regularization, namely dimensional reduction, 
was introduced in Ref.~[\ref{Sie79}] in order to explicitly preserve 
supersymmetric Ward identities in gauge theories.
This dimensional regularization prescription consists of continuing the 
virtual momenta of loop integrals into $d$ dimensions while 
keeping all (internal and external) polarizations in four dimensions. Although 
the operational definition of dimensional reduction is not complete in
some higher-loop computations [\ref{Sie80}], it has been
explicitly checked [\ref{Cap80}] that gauge and supersymmetry invariance
is maintained up to two-loop order.
In practical one-loop
calculations this prescription works like conventional dimensional 
regularization apart from a subtle (and important) point: one has
to distinguish between four-dimensional metric tensors coming from 
the Lorentz algebra of the spin indices and  
$d$-dimensional metric tensors arising from momentum integrals
with more than one loop momentum in the numerator. As for the momenta of 
the external partons one can keep them in $d$ dimensions (as originally proposed
and used, for instance, in Ref.~[\ref{susyap}]) or in four dimensions
[\ref{KST2to2}]. The latter option, in particular, is systematically used in
the string-theory inspired techniques [\ref{string}] for computing QCD one-loop
amplitudes [\ref{BDK96},\ref{BDK5g}--\ref{BDK2q3g}]: it leads to extreme 
simplifications of unnecessarily cumbersome expressions arising, for instance,
in conventional dimensional regularization.

All these various dimensional regularization prescriptions work as ultraviolet
regulators. As for gauge invariance, to our understanding the key point is
that, in the procedure of analytic continuation, there are only two
possibilites for choosing the number of external polarizations. One can set
them either to their four-dimensional value or to the number of
polarizations inside the loop. This prevents the propagation of spurious
(additional) degrees of freedom from the  loop to external states.
When we regularize ultraviolet 
singularities, these different prescriptions simply lead to different 
renormalization factors in off-shell Green functions and, possibly, to a 
perturbative shift in the definition of the renormalized running coupling.

As soon as the same prescriptions are used to regularize also soft and collinear
divergences, unitarity has to be carefully considered in the evaluation of the
on-shell matrix elements, as first pointed out in Ref.~[\ref{KST2to2}].  

In summary, we are going to discuss dimensional regularization prescriptions
that at the level of one-loop amplitudes are defined as follows. The parton 
momenta in the loop are $d$-dimensional, while the external momenta are either
$d$-dimensional (conventional dimensional regularization and dimensional
reduction as in [\ref{susyap}]) or four-dimensional ('t~Hooft and Veltman
and dimensional reduction as in [\ref{BDK96},\ref{KST2to2}]). Correspondingly,
the number $n_s(g)$ of gluon polarizations in external states is either $d-2$
or 2. The number
$h_g$ of gluon helicity states in the loop is analytically continued to 
$d$ dimensions, as in conventional dimensional regularization and in the 
't~Hooft--Veltman prescription:
\beq
\label{hgcdr}
h_g^\CDR=h_g^\HV=d-2 = 2 - 2 \ep \;\;,
\eeq
or kept fixed in four dimensions, as in dimensional reduction:
\beq
\label{hgdr}
h_g^\DR=2 \;\;.
\eeq
The number $h_q$ of massless-quark polarization states in the loop is 2 
(as in all the regularization prescriptions discussed above) or arbitrarily
continued to $d$ dimensions by $2 h_q= {\rm Tr} \,{\ID} = 4 + {\cal O}(\ep)$, 
${\rm Tr} \,{\ID}$ being the dimensionality of the spinor space
(the definition ${\rm Tr} \,{\ID}=4 - 4\ep$ was used in Ref.~[\ref{AEM}]). 
This freedom in defining $h_q$ does not simplify any 
practical calculations and is correctly referred to as harmless in any textbook
that introduces dimensional continuation as an ultraviolet regularization.
In order to emphasize in the simplest way the differences between
ultraviolet and infrared regulators, however, we also consider a toy scheme
that is identical to conventional dimensional regularization, but
with\footnote{It might seem that this is the most natural scheme, since it
corresponds to a theory in which every quantity is analytically
continued to the same number of dimensions, $d$.}
$h_q = 2-2\ep$.  We will see that this leads to effects that, although
trivial to keep track of, are not harmless.

We summarize these definitions in Table~\ref{sumtab}.
\begin{table}[th]
  \begin{center}
    \footnotesize
    \begin{tabular}{|l|c|c|c|c|c|}
      \cline{2-6}
      \multicolumn{1}{c|}{}
      & Conventional   & 't~Hooft-- & Original    & Modern      & Toy    \\
      \multicolumn{1}{c|}{}
      & dimensional    & Veltman    & dimensional & dimensional & scheme \\
      \multicolumn{1}{c|}{}
      & regularization &            & reduction   & reduction   &        \\
      \cline{2-6}
      \multicolumn{1}{c|}{}
      & CDR            & HV         &             & DR          & toy    \\
      \hline
      Number of internal dimensions
      & $d$            & $d$        & $d$         & $d$         & $d$    \\
      Number of external dimensions
      & $d$            & 4          & $d$         & 4           & $d$    \\
      Number of internal gluons, $h_g$
      & $d-2$          & $d-2$      & 2           & 2           & $d-2$  \\
      Number of external gluons, $n_s(g)$
      & $d-2$          & 2          & 2           & 2           & $d-2$  \\
      Number of internal quarks, $h_q$
      & 2              & 2          & 2           & 2           & $d-2$  \\
      Number of external quarks, $n_s(q)$
      & 2              & 2          & 2           & 2           & $d-2$  \\
      \hline
    \end{tabular}
  \end{center}
  \caption{Definitions of various regularization prescriptions of one-loop
  amplitudes referred to in the text.}
  \label{sumtab}
\end{table}

The implementation of these prescriptions in the master formula (\ref{master})
is straightforward. Following the unitarization recipe in Sect.~\ref{dipsec}, 
the tree-level matrix element $\cmrs(\{p\})$ has to be evaluated considering
the partons $\{p\}$ like the external ones in the loop amplitude. The dipole 
phase 
space is always $d$-dimensional. The dipole factors are obtained from
the emission probabilities of soft and collinear partons. However, 
soft emission, being gauge invariant and independent of the spin of the 
radiating parton, is insensitive to the treatment of the spin polarizations. 
Eventually, in order to relate different dimensional regularization 
prescriptions, we simply have to compute the corresponding   
\AP splitting functions for collinear emission.

\subsection[]{Altarelli--Parisi splitting functions in various dimensional
\linebreak
regularization schemes}
\label{apsec}

Let us consider the (time-like) splitting of a massless parton with flavour 
$a$ into two massless partons with flavours $b$ and $c$ and momenta $p_b$ and 
$p_c$. 
The collinear limit $\kt \to 0$ is precisely defined by introducing
the following parametrization of the parton momenta
\beq
\label{clim}
p_b^\mu = z p^\mu + \kt^\mu - \frac{\kt^2}{z} \frac{n^\mu}{2 p\cdot n},
\qquad p_c^\mu =
(1-z) p^\mu - \kt^\mu - \frac{\kt^2}{1-z} \frac{n^\mu}{2 p\cdot n} \;\;,
\eeq
where the light-like ($p^2=0$) vector $p^\mu$ denotes the
collinear direction and $n^\mu$ is an auxiliary light-like vector that
is necessary to specify the transverse component $\kt$ 
($\kt^2<0 , \,\kt p = \kt n = 0$) or, equivalently, how the collinear direction
is approached. 

The probability of the splitting process $a \to b + c$ (summed over colours and
spins of $b$ and $c$)
in the collinear limit 
is proportional to the Altarelli--Parisi splitting function $\P_{ab}(z,\kt)$.
Since the collinear partons $b$ and $c$ have to be treated like the partons
inside loop amplitudes, this splitting function is RS dependent. Using 
dimensional regularization the momenta in Eq.~(\ref{clim}) are always 
$d$-dimensional, but the number of polarization states of the partons $b$ and 
$c$ still depends on the detailed regularization prescription.
Moreover, the function $\P_{ab}(z,\kt)$ depends not only on the 
longitudinal-momentum
fraction $z$ involved in the splitting process (\ref{clim}) but also on
the transverse momentum $\kt$ and on the helicity of the parent parton $a$
(this parton is treated like the external partons in loop amplitudes).
More precisely, $\P_{ab}$ is a matrix acting on the spin indices of the parton 
$a$. The calculation of the polarized splitting functions is straightforward
[\ref{AP}]. We find: 
\beeq
\label{Pqq}
&&\la s|\P_{qq}(z,\kt)|s'\ra = \la s|\P_{{\bar q}{\bar q}}(z,\kt)|s'\ra =
\delta_{ss'}
\,C_F\left[\frac{2z}{1-z}+\frac{1}{2}\,\hg (1-z)\right] \;,
\\ \vspaceinarray
\label{Pqg}
&&\la s|\P_{qg}(z,\kt)|s'\ra = \la s|\P_{{\bar q}g}(z,\kt)|s'\ra =
\delta_{ss'} C_F\left[\frac{2(1-z)}{z}+\frac{1}{2}\,\hg \,z\right] \;,
\\ \vspaceinarray
\label{Pgq}
&&\la \mu|\P_{gq}(z,\kt)|\nu\ra = \la \mu|\P_{g{\bar q}}(z,\kt)|\nu\ra =
T_R \,\frac{h_q}{2}
\left[-g^{\mu\nu} + 4z(1-z)\frac{\kt^\mu\kt^\nu}{\kt^2}\right],
\\ \vspaceinarray
\label{Pgg}
&&\la \mu|\P_{gg}(z,\kt)|\nu\ra = 
2C_A \left[-g^{\mu\nu}\left(\frac{z}{1-z}+\frac{1-z}{z}\right)
- \hg z(1-z)\frac{\kt^\mu\kt^\nu}{\kt^2}\right],
\eeeq
where the spin indices of the parent parton $a$ have been denoted by $s,s'$ 
if $a$ is a fermion and by the Lorentz indices $\mu,\nu$ if $a$ is a gluon.

The scheme dependence related to the definition of $h_q$ only 
affects the Altarelli--Parisi function in Eq.~(\ref{Pgq}) for the splitting
process $g \to q + {\bar q}$. For all the other prescriptions discussed
in Sect.~\ref{rssec}, the scheme dependence is entirely parametrized by the 
number $\hg$ of gluon helicity states in the loop. This dependence consistently
vanishes for $\ep \to 0$, i.e.\ if the regularization is removed. In particular,
because of Eq.~(\ref{hgcdr}), we can immediately conclude that for the 
't~Hooft--Veltman scheme the insertion operator ${\bom I}^\RS(\{p\},\ep)$
entering into the master formula (\ref{master}) exactly coincides with
that in Eq.~(\ref{iee}) for the case of conventional dimensional regularization.

Owing to the helicity conservation in the quark--gluon vector coupling, 
the quark splitting functions in Eqs.~(\ref{Pqq},\ref{Pqg}) are 
diagonal in the spin indices. The gluon splitting functions in 
Eqs.~(\ref{Pgq},\ref{Pgg}) turn out to be diagonal after integration over the
dipole phase space. As a matter of fact, this integration involves the azimuthal 
average of the $d-2$ transverse components, which gives\footnote{In the case of 
conventional dimensional regularization the azimuthal average coincides with the
average over the polarizations of the parent gluon. In other dimensional
regularization prescriptions, however, the two averages are different,
due to the mismatch between the number, $d-2$, of transverse 
components and the number of helicity states of the external gluons.}: 
\beq
- \left\langle \;\frac{\kt^\mu\kt^\nu}{\kt^2} \;\right\rangle_{\varphi} =
\frac{1}{d-2}\left(-g^{\mu\nu} + \frac{p^\mu n^\nu+n^\mu p^\nu}{pn}\right).
\eeq
Inserting this expression into the tree-level matrix element as in 
Eq.~(\ref{insop}), the longitudinal terms proportional to $p^\mu$ and $p^\nu$
give vanishing contributions because of the Ward identity (gauge invariance)
$p^\mu \cm_\mu^\RS = 0$ and only the spin-diagonal term $-g^{\mu\nu}$ survives.

In conclusion, in any dimensional-regularization scheme the insertion operator
${\bom I}^\RS(\{p\},\ep)$ is diagonal in the helicity space and has the same 
expression as the insertion operator for conventional dimensional regularization
in Eq.~(\ref{iee}):
\beq
\label{irs}
{\bom I}^\RS(\{p\},\ep) = -
\frac{\as}{2\pi}
\frac{1}{\Gamma(1-\ep)} \sum_I \;\frac{1}{{\bom T}_{I}^2} \;{\cal
V}_I^\RS(\ep)
\; \sum_{J \neq I} {\bom T}_I \cdot {\bom T}_J
\; \left( \frac{4\pi \mu^2}{2 p_I\cdot p_J} \right)^{\ep} \;\;.
\eeq
The RS dependence is embodied in the $\ep$-expansion of the flavour functions
${\cal V}_I^\RS(\ep)$, whose coefficients are obtained by the $d$-dimensional
integration of the azimuthally averaged splitting functions
$\la {\hat P}_{ab}^\RS(z;\ep)\ra$:
\beeq
\label{Pqqav}
&&\la \P_{qq}(z;\ep) \ra = \la \P_{{\bar q}{\bar q}}(z;\ep) \ra =
C_F\left[\frac{2z}{1-z}+\frac{1}{2}\,\hg (1-z)\right] \;,
\\ \vspaceinarray
\label{Pqgav}
&&\la \P_{qg}(z;\ep) \ra = \la \P_{{\bar q}g}(z;\ep) \ra =
C_F\left[\frac{2(1-z)}{z}+\frac{1}{2}\,\hg \,z\right] \;,
\\ \vspaceinarray
\label{Pgqav}
&&\la \P_{gq}(z;\ep) \ra = \la \P_{g{\bar q}}(z;\ep) \ra =
T_R \,\frac{h_q}{2}
\left[ 1 - \frac{4}{d-2} \, z(1-z) \right] \;,
\\ \vspaceinarray
\label{Pggav}
&&\la \P_{gg}(z;\ep) \ra = 
2C_A \left[ \frac{z}{1-z}+\frac{1-z}{z} 
+ \frac{\hg}{d-2} \, z(1-z)  \right] \;.
\eeeq

Equations~(\ref{Pqqav}--\ref{Pggav}) differ by terms of order $\ep$ with 
respect to their well-known versions [\ref{ERT}] in conventional dimensional 
regularization $(h_g=d-2,$ $h_q=2)$. According to the notation in
Sect.~\ref{cdrsec}, we drop the label R.S. in conventional dimensional
regularization and parametrize the deviation from this result by the residual
scheme-dependent function ${\cal P}^\RS_{ab}(z)$ as follows:
\beq
\label{calpeq}
\la \P_{ab}(z;\ep) \ra = \la {\hat P}_{ab}(z;\ep) \ra
+ 2 \ep \;{\cal P}^\RS_{ab}(z) + {\cal O}(\ep^2) \;\;.
\eeq

Obviously, all the functions ${\cal P}^\CDR_{ab}(z)$ as well as 
${\cal P}^\HV_{ab}(z)$ vanish. 

In the toy scheme, which differs from conventional dimensional regularization
simply because of the dimensionality of the gamma matrices in quark loops, the 
only non-vanishing ${\cal P}$-functions are:
\beq
\label{cptoy}
{\cal P}^{{\rm toy}}_{gq}(z) = {\cal P}^{{\rm toy}}_{g{\bar q}}(z) =
- \;\frac{1}{2} \;T_R \;\left[ z^2 + (1-z)^2 \right] \;\;.
\eeq

For dimensional reduction we have:
\beeq
\label{PDRq}
&&{\cal P}^\DR_{qq}(z)= {\cal P}^\DR_{{\bar q}{\bar q}}(z) =
C_F \;\frac{1-z}{2} \;,
\qquad\quad
{\cal P}^\DR_{qg}(z)= {\cal P}^\DR_{{\bar q}g}(z) =
C_F \;\frac{z}{2} \;,
\\ \vspaceinarray
\label{PDRg}
&&{\cal P}^\DR_{gg}(z)=C_A \,z (1-z),
\qquad\qquad\qquad\quad
{\cal P}^\DR_{gq}(z)= {\cal P}^\DR_{g{\bar q}}(z) = 0 \;.
\eeeq
In this case, some other comments are in order.
Since $h_g^\DR=2$, the quark splitting functions
$\la {\hat P}_{qq}^\DR(z;\ep) \ra$ and $\la {\hat P}_{qg}^\DR(z;\ep) \ra$
in Eqs.~(\ref{Pqqav},\ref{Pqgav}) are actually independent of $\ep$,
while the gluon splitting functions $\la {\hat P}_{gq}^\DR(z;\ep) \ra$
and $\la {\hat P}_{gg}^\DR(z;\ep) \ra$ are not. In spite of having used
only four-dimensional polarizations, the $\ep$-dependence enters into
the gluon splitting functions through azimuthal correlations, whose average
has to be carried out in $d-2$ transverse dimensions. 

Note also that the difference between the splitting functions in conventional
dimensional regularization and those in dimensional reduction can easily be 
understood on field-theoretical basis. As discussed in detail in 
Ref.~[\ref{Cap80}], the relationship with conventional dimensional 
regularization can be investigated by splitting the four-dimensional
gauge field of the dimensionally-reduced Lagrangian ${\cal L}^\DR$
into a $d$-dimensional vector $A_\mu$ plus $4-d=2\ep$ additional components.
This leads to the decomposition ${\cal L}^\DR = {\cal L}^\CDR + 
{\cal L}^{(\ep)}$,
where ${\cal L}^\CDR$ is the term that involves only the $d$-dimensional
vector field and coincides with the customary Lagrangian in conventional 
dimensional regularization. The additional $2\ep$ components enter into 
${\cal L}^{(\ep)}$ and behave as scalar gluons interacting with the fermion
and vector fields. The ultraviolet role of these $2\ep$ scalars was pointed out
in Ref.~[\ref{Cap80}]. For instance, the exact relation between the customary 
$\overline {\rm MS}$ coupling and the minimally-subtracted renormalized coupling
in dimensional reduction can be derived by a simple
calculation as the effect of the additional $2\ep$ scalars.
Equation~(\ref{calpeq}) displays the infrared role of these $2\ep$
`ghost states'. Indeed, the functions ${\cal P}^\DR_{qq}(z)$, 
${\cal P}^\DR_{qg}(z)$ and ${\cal P}^\DR_{gg}(z)$ in 
Eqs.~(\ref{PDRq},\ref{PDRg})
are exactly the Altarelli--Parisi probabilities for the splitting processes
$q \to q(z) + \phi(1-z)$, $q \to \phi(z) + q(1-z)$ and $g \to  \phi(z) +
\phi(1-z)$, where $\phi$ is a scalar gluon.

Finally, we recall that setting $C_F=T_R=C_A$ in a tree-level QCD calculation
leads to recovering the results in $N=1$ supersymmetric Yang--Mills theory 
[\ref{susyYM}]. The quark is replaced by the gluino $\tilde{g}$ and the
corresponding {\em four-dimensional} Altarelli--Parisi probabilities
fulfil a well-known supersymmetric Ward identity, namely
$P_{gg}(z) + P_{g\tilde{g}}(z) = P_{\tilde{g}g}(z) + P_{\tilde{g}\tilde{g}}(z)$.
As for the $d$-dimensional Altarelli--Parisi splitting functions in 
Eqs.~(\ref{Pqqav}--\ref{Pggav}), their version in conventional
dimensional regularization violates a similar Ward identity.
Using dimensional reduction, instead, we have
\beq\label{susyidentity}
\la {\hat P}_{gg}^\DR(z;\ep) \ra + \la {\hat P}_{g\tilde{g}}^\DR(z;\ep) \ra
= \la {\hat P}_{\tilde{g}g}^\DR(z;\ep) \ra + 
\la {\hat P}_{\tilde{g}\tilde{g}}^\DR(z;\ep) \ra \;\;,
\eeq
thus leading to equal decay probabilities for the two supersymmetric partners
in {\em any\ } number $d$ of space-time dimensions. This result is expected
for a supersymmetric regularization.

We note that in the toy scheme, in which the number of quark states is
analytically continued to be the same as the number of gluon states,
$h_q=h_g=2-2\ep,$ the 
same Ward identity is also recovered:
\beq
\la {\hat P}_{gg}^{{\rm toy}}(z;\ep) \ra +
\la {\hat P}_{g\tilde{g}}^{{\rm toy}}(z;\ep) \ra
= \la {\hat P}_{\tilde{g}g}^{{\rm toy}}(z;\ep) \ra + 
\la {\hat P}_{\tilde{g}\tilde{g}}^{{\rm toy}}(z;\ep) \ra \;\;.
\eeq

\subsection[]{The master function $\cf$ and RS independence\\of the cross
sections}

The terms of order $\ep$, which arise in the splitting functions computed 
with different regularization prescriptions, combine with $1/\ep$-poles coming
from collinear singularities in the integrals of the dipole factors and thus
provide the insertion operator ${\bom I}^\RS$ in Eq.~(\ref{master})
with finite corrections.

More precisely, as discussed in the previous subsection, 
the RS dependence of the insertion
operator ${\bom I}^\RS$ in dimensional regularization is entirely taken into
account by the flavour functions ${\cal V}_I^\RS(\ep)$. The coefficients of
the $\ep$-expansion of the functions ${\cal V}_I^\RS(\ep)$ are related to
$d$-dimensional integrals of the splitting functions, as in Eq.~(\ref{coef}).
Different regularization prescriptions lead to contributions
of the order of $\ep$ to the azimuthally-averaged splitting functions
$\la \P_{ab}(z;\ep) \ra$ and, in turn, to Eq.~(\ref{coef}). 
These contributions produce a scheme-dependent shift in the constants 
$K_a$ on the right-hand side of Eq.~(\ref{coef}) and, hence, non-singular
corrections in ${\cal V}_I^\RS(\ep)$. Parametrizing these corrections
as differences with respect to the flavour functions  ${\cal V}_I(\ep)$ in
conventional dimensional regularization, we have
\beq
\label{cvrs}
{\cal V}_I^\RS(\ep) =
{\cal V}_I(\ep) - \gtil_I^\RS + {\cal O}(\ep) \;,
\eeq
where the scheme-dependent coefficients $\gtil_I^\RS$ are obtained by inserting
Eq.~(\ref{calpeq}) into Eq.~(\ref{coef}):
\beq
\label{gtilgen}
\gtil_a^\RS 
= \sum_b \int_0^1 dz \;{\cal P}^\RS_{ab}(z) \;\;.
\eeq

Using Eq.~(\ref{gtilgen}) one can straightforwardly evaluate $\gtil_I^\RS$ in 
different RSs. In particular, we have
\beq
\gtil_I^\HV = 0 \;\;,
\eeq
and, using Eq.~(\ref{cptoy}) and Eqs.~(\ref{PDRq},\ref{PDRg}), we respectively
find
\beeq
\label{gtoy}
&&\gtil_q^{{\rm toy}} = \gtil_{{\bar q}}^{{\rm toy}} = 0 \;,
\qquad\qquad \gtil_g^{{\rm toy}}= - \frac{2}{3}\,T_R N_f \;\;, \\
\label{gdr}
&&\gtil_q^\DR = \gtil_{{\bar q}}^\DR = \frac{1}{2}\,C_F \;,
\qquad \gtil_g^\DR = \frac{1}{6}\,C_A \;\;.
\eeeq

Equation~(\ref{cvrs}) can be inserted in Eq.~(\ref{irs}) to derive a simple
expression for the RS dependence of the operator ${\bom I}^\RS$ 
in dimensional regularization. As a matter of fact, neglecting 
${\cal O}(\ep)$ corrections and using colour-charge conservation,
$\sum_{J \neq I} {\bom T}_J = - {\bom T}_I$ (the
insertion operator acts onto the vector $|1,\dots,n\ra_\RS$, which is
a colour-singlet state and, hence, fulfils the property 
$\sum_J {\bom T}_J |1,\dots,n\ra_\RS = 0$), we obtain
\beq
{\bom I}^\RS(\{p\},\ep) = {\bom I}(\{p\},\ep) - \frac{\as}{2\pi}
\sum_I {\tilde \gamma}_I^\RS + {\cal O}(\ep) \;\;.
\eeq 
In terms of the master function in Eq.~(\ref{master}), we can write:
\beq
\label{calf}
\cf(\{p\}) = |\cmrs(\{p\})|^2_{{\rm 1-loop}} \;- \frac{\as}{2\pi}
\;|\cmrs(\{p\})|^2 \;\sum_I {\tilde \gamma}_I^\RS \;+
\;|\cmrs(\{p\})|^2 \otimes 
{\bom I}(\{p\},\ep) \;\;.
\eeq
As usual, ${\bom I}(\{p\},\ep)$ refers to conventional dimensional 
regularization and is given in Eq.~(\ref{iee}).

Let us first comment on the result in Eq.~(\ref{calf}) in the context of 
the toy scheme in which $h_q = 2-2\ep$. The renormalization of the
ultraviolet divergences
in the off-shell loop amplitudes is performed exactly as in conventional
dimensional regularization. The only difference regards the definition of the
running coupling. Using modified minimal subtraction, one introduces
a renormalized coupling $\alpha_S^{{\rm toy}}$ that is related to the
customary $\overline {\rm MS}$ coupling by $\alpha_S^{{\rm toy}} =
\as [ 1 - \as T_R N_f/3\pi + {\cal O}(\as^2)]$. This is 
the harmless dependence on the ultraviolet regularization: the only
physical consequence in an overall redefinition of the coupling
constant. As soon as we
consider the renormalized on-shell one-loop amplitude on the right-hand side
of Eq.~(\ref{calf}), we encounter soft and collinear divergences that are
cancelled by the insertion operator ${\bom I}(\{p\},\ep)$, computed in 
conventional dimensional regularization. However, according to our calculation,
the remaining contribution has still to be corrected by the finite 
$\gtil_I^{{\rm toy}}$-terms in order to give a function $\cf$ consistent with
unitarity and, hence, RS-independent cross sections. This additional unitarity
correction does depend on the process through the tree-level matrix element
and its flavour topology $(\gtil_q^{{\rm toy}} \neq \gtil_g^{{\rm toy}}$).

In the context of the 't~Hooft--Veltman and dimensional-reduction prescriptions,
the scheme dependence of the one-loop QCD amplitudes was studied in 
Ref.~[\ref{KST2to2}]. On the basis of the explicit evaluation of the one-loop
corrections to all $2\to 2$ QCD subprocesses, transition rules to relate
these schemes to conventional dimensional regularization were derived (and
argued to be universal).
They were confirmed by the calculation in Ref.~[\ref{BDK2q3g}].
We can use these rules as a consistency check of our calculation and, 
in general, of our unitarization recipe. When inserted in Eq.~(\ref{calf}),
the coefficients $\gtil_I^\DR$ in Eq.~(\ref{gdr}) provide the master
function $\cf$ with the contributions that are necessary to exactly cancel
the RS dependence of the one-loop amplitudes $|\cmrs(\{p\})|^2_{{\rm 1-loop}}$
as computed in Ref.~[\ref{KST2to2}].

We can turn this argument round. Writing Eq.~(\ref{calf}) as follows:
\beq
\label{oneloopdr}
|\cmrs(\{p\})|^2_{{\rm 1-loop}} = - |\cmrs(\{p\})|^2 \otimes 
{\bom I}(\{p\},\ep)
+ \frac{\as}{2\pi} \;|\cmrs(\{p\})|^2 \;\sum_I {\tilde \gamma}_I^\RS
+ \cf(\{p\}) \;\;,
\eeq
and knowing that the function $\cf$ is RS independent, we can compute the
scheme dependence of the one-loop amplitudes $|\cmrs(\{p\})|^2_{{\rm 1-loop}}$
from our calculation of the coefficients ${\tilde \gamma}_I^\RS$. This is a
way to rederive the transition rules found in Ref.~[\ref{KST2to2}]. 
Since the ${\tilde \gamma}_I^\RS$'s are obtained from the integral 
of the dipole factors, this derivation is universal (the dipole factors do not
depend on any specific QCD amplitude) and essentially involve a tree-level
calculation (the emitted partons in the dipole are on shell) instead of a 
loop one.

Concluding this section on dimensional regularization, we should add a marginal
comment. The RS dependence of the \AP splitting functions may, in principle,
also affect the factorization-scheme dependence of the NLO partonic cross
sections if the collinear counterterm in Eq.~(\ref{sNLO}) is not defined
accordingly (for instance, if, independently of the regularization procedure,
one uses the same minimally-subtracted expression for
$d\sigma_C$). In the notation of Eq.~(\ref{sNLO4}), this dependence is embodied
in the contribution ${\hat \sigma}^{NLO\,\{m\}}$ and, in particular,
within the dipole algorithm, it is explicitly taken into account by the  
colour-charge insertion operators $\bom K$ and $\bom H$
of Ref.~[\ref{CSdipole}].
If one is interested in changing factorization scheme, 
${\hat \sigma}^{NLO\,\{m\}}$ (e.g.\ the operators  $\bom K$ and $\bom H$)
and the parton distributions $f$ and $d$ in Eq.~(\ref{hadxs}) have to be varied
consistently. We think that pursuing the study of different
$d$-dimensional factorization prescriptions has no particular 
relevance, in practice.

\section{Conclusion}
\label{conc}

In this paper we have considered some unitarity issues related to the 
regularization of unphysical soft and collinear divergences in perturbative QCD
computations. Our analysis has been performed in the framework of the
dipole formalism. It allows one to work out a general discussion while
providing an explicit implementation of unitarity constraints
in the calculation of QCD cross sections at NLO.

In Sect.~\ref{dipsec} we have shown that
the regularization-scheme independence of physical observables is
controlled by the following master function
\beq
\label{masterc}
\cf(\{p\}) = |\cmrs(\{p\})|^2_{{\rm 1-loop}} \;+  
\;|\cmrs(\{p\})|^2 \otimes 
{\bom I}^\RS(\{p\}) \;\;,
\eeq
which involves the one-loop and tree-level matrix elements and the
integral ${\bom I}^\RS(\{p\})$ of the dipole factors. The latter are obtained
by the factorization formulae for soft and collinear emission. Using a 
unitary regularization scheme, infrared singularities as well as finite
scheme-dependent remainders consistently cancel by combining the several 
contributions on the right-hand side of Eq.~(\ref{masterc}).

This result can also be used to relate one-loop amplitudes 
in different regularization schemes without explicitly carrying out any
loop calculations. Turning Eq.~(\ref{masterc}) round,
\beq
\label{masterinv}
|\cmrs(\{p\})|^2_{{\rm 1-loop}} = 
- |\cmrs(\{p\})|^2 \otimes {\bom I}^\RS(\{p\}) \;+ \cf(\{p\}) \;\;,
\eeq
and exploiting the scheme independence of $\cf(\{p\})$, one can obtain
$|\cmrs(\{p\})|^2_{{\rm 1-loop}}$ in a different scheme by simply evaluating
the difference of the corresponding dipole factors.

More generally, our explicit construction of Eq.~(\ref{masterc}) allows one
to compute physical cross sections also using regularization prescriptions 
that are not manifestly unitary. As long as the regularization procedure
is consistently defined at the level of one-loop amplitudes, we can give
a recipe (see Sect.~\ref{dipsec}) to calculate the dipole-factor 
contributions that are necessary to guarantee unitarity. From the regularized
virtual corrections one can thus extract the essential physical information, 
that is, the regularization-scheme independent function $\cf(\{p\})$.

In order to make our general discussion more definite we have considered, in
Sect.~\ref{dimregsec}, the case of dimensional-regularization prescriptions.
This analysis required the explicit calculation of the $d$-dimensional
\AP splitting functions in various regularization schemes.
Our results for these functions differ from previously published ones in
the 't~Hooft--Veltman and dimensional reduction schemes
[\ref{GGjets},\ref{KST2to2}]. These differences emerge in the
$\ep$-dependent terms as can be seen by comparing our formul\ae\
(\ref{calpeq}--\ref{PDRg}) with the corresponding results in
Refs.\ [\ref{GGjets},\ref{KST2to2}]. The differences come about because
in these references the average over polarizations of the parent parton
was taken, instead of the azimuthal average.  These are not the
same in the case of gluon splitting functions as discussed in
Sect.~\ref{apsec}.
However, this slight error has no practical consequence on the main results of
Refs.\ [\ref{GGjets},\ref{KST2to2}] 
because in those references the relationship between
splitting functions and RS dependence was not exploited
(in Ref.~[\ref{KST2to2}], the splitting functions were used to relate the
definition of different factorization schemes).
On the contrary, we make full use of this relationship, allowing us to
derive the unitarity corrections needed to relate different
dimensional-regularization schemes (see Eq.~(\ref{oneloopdr})), without
having to make any loop calculations.
Our results are in full agreement with those obtained in Ref.~[\ref{KST2to2}].
We are able to give a probabilistic interpretation (see
Eq.~(\ref{gtilgen})) of the coefficients~$\gtil_I^\RS$.

The method can also be applied to other regularization procedures, such as,
for instance, massive regularization of the loop integrals. 
Many higher-order calculations in QED have been carried out using
this regularization scheme. Dimensional regularization is certainly 
preferred in massless QCD and
its computational advantages have also been exploited in recent QED 
achievements [\ref{g-2}]. 
Our unitarization technique and the extension of the dipole formalism to 
massive partons [\ref{CS}] can be convenient in order to
use known QED results in QCD applications and to combine them 
with new QED calculations in dimensional regularization.

Systematic QCD calculations at next-to-next-to-leading order (NNLO) for jet
observables will become feasible only when efficient techniques for evaluating
two-loop matrix elements will be set up. At that point the unitarity issue
discussed in this paper will show up again.
Our results may eventually be very useful to tackle this issue, 
provided the validity of the
dipole formalism is extended to such a level of accuracy.

In concluding, we would like to point out another feature of 
Eq.~(\ref{masterc}). The operator ${\bom I}^\RS(\{p\})$ is obtained by
integrating the dipole factors and, correspondingly, 
$|\cmrs(\{p\})|^2_{{\rm 1-loop}}$ is the result of the integration over the loop
momentum:
\beq
\label{masterint}
\cf(\{p\}) =  \int_{{\rm loop}} d|\cmrs(\{p\})|^2_{{\rm 1-loop}}
+ \sum_{{\rm dipoles}} \;\int_1 \;|\cmrs(\{p\})|^2 \otimes 
dV_{{\rm dipole}}^\RS \;\;.
\eeq
It is conceivable that one may find a way of combining the two integrands
such that the dipole factors act as a local counterterm for the loop
integral. Achieving this, one could avoid the introduction of any soft
and collinear regularization and the ensuing unitarity problems. Most 
importantly, one would be able to carry out NLO calculations of
physical cross sections by the sole use of numerical methods, without
any analytical calculation of one-loop amplitudes.

\bigskip 
\noindent{\bf Acknowledgements. \ } 
This research is supported in part by EEC Programme {\it Human Capital
and Mobility}, Network {\it Physics at High Energy Colliders},
contracts CHRX-CT93-0357 (DG 12 COMA) and PECO ERBCIPDCT 94 0613, 
the Hungarian Science Foundation grant OTKA T-016613 and the Research Group in
Physics of the Hungarian Academy of Sciences, Debrecen.

\section*{References}

\def\ac#1#2#3{Acta Phys.\ Polon.\ #1 (19#3) #2}
\def\ap#1#2#3{Ann.\ Phys.\ (NY) #1 (19#3) #2}
\def\ar#1#2#3{Ann.\ Rev.\ Nucl.\ Part.\ Sci.\ #1 (19#3) #2}
\def\cpc#1#2#3{Computer Phys.\ Comm.\ #1 (19#3) #2}
\def\ib#1#2#3{ibid.\ #1 (19#3) #2}
\def\np#1#2#3{Nucl.\ Phys.\ B#1 (19#3) #2}
\def\pl#1#2#3{Phys.\ Lett.\ #1B (19#3) #2}
\def\pr#1#2#3{Phys.\ Rev.\ D #1 (19#3) #2}
\def\prep#1#2#3{Phys.\ Rep.\ #1 (19#3) #2}
\def\prl#1#2#3{Phys.\ Rev.\ Lett.\ #1 (19#3) #2}
\def\rmp#1#2#3{Rev.\ Mod.\ Phys.\ #1 (19#3) #2}
\def\sj#1#2#3{Sov.\ J.\ Nucl.\ Phys.\ #1 (19#3) #2}
\def\zp#1#2#3{Zeit.\ Phys.\ C#1 (19#3) #2}

\begin{enumerate}

\item \label{BDK96}
Z.\ Bern, L.\ Dixon and D.A.\ Kosower, preprint SLAC-PUB-7111
(hep-ph/9602280) and references therein.

\item \label{GGjets}
W.T. Giele and E.W.N. Glover, \pr{46}{1980}{92};
W.T. Giele, E.W.N. Glover and D.A. Kosower, \np{403}{633}{93}.

\item \label{subm}
Z.\ Kunszt and D.E.\ Soper, \pr{46}{192}{92};
S.\ Frixione, M.L.\ Mangano, P.\ Nason and G.\ Ridolfi, \np{412}{225}{94}.

\item \label{FKSjets}
S.\ Frixione, Z.\ Kunszt and A.\ Signer, \np{467}{399}{96}.

\item \label{CSLett}
S.\ Catani and M.H.\ Seymour, \pl{378}{287}{96};
preprint CERN-TH/96-181 (hep-ph/9607318).

\item \label{CSdipole}
S.\ Catani and M.H.\ Seymour, \np{485}{291}{97}.

\item \label{NTjets}
Z. Nagy and Z. Tr\'ocs\'anyi, preprint KLTE-DTP/96-1 (hep-ph/9610498).

\item \label{BDK5g}
Z.\ Bern, L.\ Dixon and D.A. Kosower, \prl{70}{2677}{93}.

\item \label{KST4q1g}
Z.\ Kunszt, A.\ Signer and Z. Tr\'ocs\'anyi, \pl{336}{529}{94}.

\item \label{BDK2q3g}
Z.\ Bern, L.\ Dixon and D.A.\ Kosower, \np{437}{259}{95}.

\item \label{pp3jetg}
Z. Tr\'ocs\'anyi, \prl{77}{2182}{96}.

\item \label{kilgore}
W.B.\ Kilgore, preprint Fermilab-Conf-96/315-T (hep-ph/9609367);
W.B.\ Kilgore and W.T. Giele, preprint Fermilab-Pub-96/358-T (hep-ph/9610433). 

\item \label{4jleading}
A. Signer and L. Dixon, preprint SLAC-PUB-7309 (hep-ph/9609460).

\item \label{4jqqqq}
E.W.N. Glover and D.J. Miller, Durham preprint DTP/96/06 (hep-ph/9609474);
Z.~Bern, L.\ Dixon, D.A.\ Kosower and S.\ Weinzierl, preprint SLAC-PUB-7316 
(hep-ph/9610370).
 
\item \label{Sie79}
W. Siegel, \pl{84}{193}{79}.

\item \label{KST2to2}
Z. Kunszt, A. Signer and Z. Tr\'ocs\'anyi, \np{411}{397}{94}.

\item \label{verm}
J.A.M.\ Vermaseren, K.J.F.\ Gaemers and S.J. Oldham, \np{187}{301}{81}.

\item \label{cant}
F.\ Bloch and A.\ Nordsieck, Phys.\ Rev.\ 52 (1937) 54;
T.\ Kinoshita, J.\ Math.\ Phys.\ 3 (1962) 650; 
T.D.\ Lee and M.\ Nauenberg, Phys.\ Rev.\ B133 (1964) 1549.

\item  \label{BCM}
See, for instance: A.\ Bassetto, M.\ Ciafaloni and G.\ Marchesini,
\prep{100}{201}{83}; Yu.L.\ Dokshitzer, V.A.\ Khoze, A.H.\ Mueller and
S.I.\ Troyan, {\em Basics of Perturbative QCD}\/, Editions Fronti\`eres,
Paris, 1991.

\item \label{ERT}
R.K. Ellis, D.A. Ross and A.E. Terrano, \np{178}{421}{81}.

\item \label{dope}
R.P. Feynman, \ac{26}{697}{63}.

\item \label{tHV}
G. 't~Hooft and M.\ Veltman, \np{44}{189}{72}.

\item \label{bollini}
G.\ Bollini and J.J.\ Giambiagi, Nuovo\ Cimento\ 12B (1972) 20;
J.F.\ Ashmore, Nuovo\ Cimento\ Lett. 4 (1972) 289;
G.M.\ Cicuta and E.\ Montaldi, Nuovo\ Cimento\ Lett. 4 (1972) 329.

\item \label{cdrir}
R.\ Gastmans and R.\ Meuldermans, \np{63}{277}{73}.

\item \label{Sie80}
W.\ Siegel, \pl{94}{37}{80};
L.V.\ Avdeev and A.A.\ Vladimirov, \np{219}{262}{83}.

\item \label{Cap80}
D.M.\ Capper, D.R.T.\ Jones and P.\ van Nieuwenhuizen, \np{167}{479}{80}.

\item \label{susyap}
I.\ Antoniadis and E.G.\ Floratos, \np{191}{217}{81}.

\item \label{string}
Z.\ Bern and D.A.\ Kosower, \np{379}{451}{92}.

\item \label{AEM}
G.\ Altarelli, R.K.\ Ellis and G.\ Martinelli, \np{157}{461}{79}.

\item \label{AP}
G. Altarelli and G. Parisi, \np{126}{298}{77}.

\item \label{susyYM}
J. Wess and B. Zumino, \np{70}{39}{74}.

\item \label{g-2}
S.\ Laporta and E.\ Remiddi, \pl{379}{283}{96}.

\item \label{CS}
S.\ Catani and M.H.\ Seymour, in preparation.

\end{enumerate}

\end{document}